\documentclass[twocolumn,pra]{revtex4}
\usepackage{amssymb}

\usepackage{graphicx}
\usepackage{dcolumn}

\begin{document}

\title{Dynamical transitions in a modulated Landau-Zener model with finite
driving fields}

\author{Wei Li}
\affiliation{Center of Theoretical Physics, College of Physical Science and
Technology, Sichuan University, Chengdu 610065, China}
\author{Li-Xiang Cen}
\email{lixiangcen@scu.edu.cn}
\affiliation{Center of Theoretical Physics, College of Physical Science and
Technology, Sichuan University, Chengdu 610065, China}

\begin{abstract}
We investigate a special time-dependent quantum model which assumes the
Landau-Zener driving form but with an overall modulation of the intensity of
the pulsing field. We demonstrate that the dynamics of the system, including
the two-level case as well as its multi-level extension, is exactly solvable
analytically. Differing from the original Landau-Zener model, the
nonadiabatic effect of the evolution in the present driving process does not
destroy the desired population transfer. As the sweep protocol employs only
the finite driving fields which tend to zero asymptotically, the cutoff
error due to the truncation of the driving pulse to the finite time interval
turns out to be negligibly small. Furthermore, we investigate the
noise effect on the driving protocol due to the dissipation of
the surrounding environment. The losses of the fidelity in the protocol
caused by both the phase damping process and the random spin flip noise are
estimated by solving numerically the corresponding master equations within
the Markovian regime.
\end{abstract}

\maketitle

\section{introduction}

Exactly solvable time-dependent quantum system attracts increasing interest
owing to its role in the design for quantum control. In particular, to model
dynamical processes or target quantum states for atomic and molecular
systems \cite{nonadt,brumer}, nonadiabatic transitions induced by time-varying
external fields are often involved and the theoretical proposal of the
driving protocol with desired dynamics is generally a prerequisite to
accomplish the corresponding
quantum tasks \cite{LZ1,LZ2,cd1,cd2,tranl,short,barn1,barn2,mess,exp1,exp2,exp3}.

Landau-Zener (LZ) model \cite{LZ1,LZ2} and its analogs, represented by the
Hamiltonian below, are the most frequently exploited proposals in the
driving protocol
\begin{equation}
H(t)=\Omega_x(t)J_x+\Omega_z(t)J_z.  \label{hamil0}
\end{equation}
Here $J_{x,z}$ denote the angular-momentum operators and $\Omega_{x,z}(t)$
account for two components of the driving field along the $x$ and $z$
axes, respectively. Owing to the explicit time dependency of $H(t)$, the
general solution to this kind of systems is highly nontrivial even for the
simplest two-level case, i.e., with the azimuthal quantum number $j=\frac12$.
For the standard LZ sweep with $\Omega_x$ being
constant and $\Omega_z(t)$ varying linearly with time, the very two-level
model is exactly solvable and the transition probability induced by the
evolution over $t\in(-\infty,\infty)$ is known well as the LZ formula \cite
{LZ1,LZ2}. Notably, the LZ model has a wide range of applications in physics
as well as in chemistry, including the LZ interferometry \cite{LZI1,LZI2,LZI3,LZI4},
the transfer of charge \cite {charge}, chemical reactions \cite{chem1,chem2},
controllable manipulation of qubit and qutrit systems \cite{manip,triq,you,qtrit},
and so on.

The so-called counter-diabatic protocol \cite{cd1,cd2} (also named as the
transitionless protocol \cite{tranl} or shortcuts to adiabaticity \cite
{short}) has been proposed to generate exact dynamical evolution which aims
at the adiabatic eigenstates, e.g., of a given Hamiltonian of form (\ref
{hamil0}). Typically, this kind of protocols exploit a reverse-engineering strategy
through introducing an auxiliary counter-diabatic driving term [e.g., an
extra time-varying field along the $y$-axis which cancels out the nonadiabatic
effect of $H(t)$] to ensure the desired evolution. We would also like to
mention another reverse-engineering algorithm proposed in Ref. \cite{barn1},
where a parametric connection is established between the the evolution operator and the
control field of the Hamiltonian.
In comparison, while the latter method is able to generate the LZ-type
protocol with two driving components formed of Eq. (\ref{hamil0}),
its applications were restricted to the two-level systems \cite{barn1,barn2,mess}.

Except for the models constructed through the mentioned
reverse-engineering methods, analytically exactly solvable time-dependent
quantum systems are relatively rare and known examples are mostly
concentrated on the two-level system, for example, the Rosen-Zener \cite
{rosen}, Allen-Eberly \cite{allen}, Demkov-Kunike \cite{demkov}, and
Bambini-Berman \cite{bamb} models. In a recent work, a tangent-pulse
driven model has been proposed \cite{tan} which is shown to be analytically
solvable not only
for the two-level case but also for the multi-level extension.
The nonadiabatic dynamics generated by the model itself can serve as a desirable
protocol for the population transfer without the need of any auxiliary
fields. While the ideal design assumes an infinite chirping
field, it is demonstrated that for an imperfect scanning process with
truncation, the cutoff error caused to the population transfer could be suppressed
to the infinity through enhancing the scanning rate of the protocol.

In this paper we propose a modulated LZ model and explore the generated
dynamics for quantum control.
In particular, we demonstrate that the model offers an alternative protocol for
the nonadiabatic population transfer which retains
the advantages previously displayed in the tangent-pulse driven model:
the nonadiabatic evolution can realize complete population transfer and no
auxiliary field is required; the model is genuinely solvable which can be
extended to the multi-level system. Furthermore, since the present protocol
employs only the fields of finite intensity, it avoids the nonrealistic
design of infinite driving assumed in the original LZ model and other
analogous schemes. Meanwhile, the cutoff
error in the protocol due to the truncation of the scanning pulse to the
finite time interval is shown to be negligibly small. To evaluate further the
feasibility of the scheme in the real systems, we investigate the noise
effect of the protocol under dissipation. We solve numerically the master
equations associated with the dephasing process and the random spin
flip process within the Markovian regime. The loss of the fidelity caused
by the detrimental influence of the noise is estimated.

The rest of the paper is organized as follows. In Sec. II we will introduce
the modulated LZ model and demonstrate that the dynamics of model
governed by the time-dependent Schr\"{o}dinger equation is exactly solvable.
We will employ the method proposed
by Lewis and Riesenfeld (LR) \cite{lewis1,lewis2} and manifest explicitly
the dynamical invariant of the model. In Sec. III we shall focus on the
dynamical transition in the model and describe the corresponding process of
nonadiabatic population transfer for the two-level case as well as for its
multi-level extension. Especially, we show that the intermediate transitions
induced by the nonadiabatic effect will not destroy the desired state transfer.
The noise effect on the fidelity of the protocol due to the dissipation of
the environment is investigated in Sec. IV. Finally, a summary of the paper
is presented in Sec. V.

\section{Description of the model and its exact solution}

The driven model considered here is described explicitly by the Hamiltonian
\begin{equation}
H(t)=\frac \eta {1+\nu ^2t^2}(J_x+\kappa \nu tJ_z),  \label{hamil1}
\end{equation}
where the amplitude $\eta $ and the sweep frequency $\nu $ are fixed
constants and the coefficient $\kappa $ relates to them via
\begin{equation}
\kappa =\sqrt{1-(\nu /\eta )^2}.
\label{cond}
\end{equation}
Here we have set $\hbar=1$ such that $\eta\equiv\eta/\hbar$ possesses the same
dimension with $\nu$.
As the model keeps the property of the original LZ model that the ratio
between the field components along the $z$ and $x$ axes $\Omega _z(t)/\Omega
_x(t)$ increases linearly with time, an overall modulation on the field
amplitude is exploited in the present sweep process. The schematic of the
scanning pulses $\Omega _{x,z}(t)$ over $t\in (-\infty ,\infty )$ is depicted in
Fig. 1. Note that the modulation here enables the model to avoid the
nonrealistic ingredient of assuming an infinite driving field in the
original model.

\begin{figure}[b]
\includegraphics[width=0.9\columnwidth]{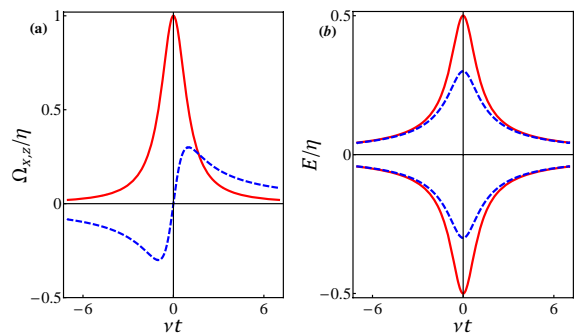}
\caption{The scanning process of the modulated Landau-Zener
model specified by Eq. (1): (a) Time dependence of the two field components
$\Omega_x(t)/\eta$ (solid line) and $\Omega_z(t)/\eta$ (dashed line) with
$\kappa=0.6$ (i.e., $\nu/\eta=0.8$). (b) The corresponding adiabatic (solid
line) and diabatic (dashed line) energy levels, $E_{\pm}^{ad}(t)$ and
$E_{\pm}(t)$ over $\eta $ of the two-level system with $\kappa \rightarrow 1$ and
$\kappa =0.6$, respectively. The levels exhibit maximal splits
at $t=0$ with $E_{\pm}^{ad}(0)/\eta=\mp 0.5$ and
$E_{\pm}(0)/\eta=\mp 0.3$.}
\end{figure}

We now show that the dynamics of the system governed by the Schr\"{o}dinger
equation $i\partial _t|\psi (t)\rangle =H(t)|\psi (t)\rangle $ is exactly
solvable. To this goal, we recall the dynamical invariant introduced by the
LR method \cite{lewis1,lewis2}. That is, a time-dependent quantum system
could be solved exactly if the system possesses a dynamical invariant, i.e.,
an observable $I(t)$ that satisfies
\begin{equation}
i\frac{\partial I(t)}{\partial t}-[H(t),I(t)]=0.  \label{lewis}
\end{equation}
The peculiar property of such an invariant is that its instantaneous
eigenvector, denoted by $|\phi _m(t)\rangle $, differs from the basic
solution to the Schr\"{o}dinger equation only by a phase factor: $|\psi
_m(t)\rangle =e^{i\Phi _m(t)}|\phi _m(t)\rangle $, in which $\Phi _m(t)$ is
expressed as
\begin{equation}
\Phi _m(t,t_0)=\int_{t_0}^t\langle \phi _m(t^{\prime })|i\frac \partial
{\partial t^{\prime }}-H(t^{\prime })|\phi _m(t^{\prime })\rangle dt^{\prime
}.  \label{LRphase}
\end{equation}
Exact analytical expression of the LR invariant
has ever been found for the time-dependent quantum system of
particular classes \cite{wang,HQC,Dang}.
Intriguingly, the above system is shown to possess the following invariant
\begin{eqnarray}
I(t) &=&\vec{\alpha}(t)\cdot \vec{J}  \nonumber \\
&=&\frac 1{\sqrt{1+\nu ^2t^2}}(\kappa J_x+\frac \nu \eta J_y+\nu tJ_z).
\label{invar}
\end{eqnarray}
It is direct to verify that the specified $\alpha _i(t)$ satisfy
\begin{eqnarray}
\dot{\alpha}_x(t) &=&-\Omega _z(t)\alpha _y(t),  \nonumber \\
\dot{\alpha}_y(t) &=&\Omega _z(t)\alpha _x(t)-\Omega _x(t)\alpha _z(t),
\nonumber \\
\dot{\alpha}_z(t) &=&\Omega _x(t)\alpha _y(t),  \label{relat}
\end{eqnarray}
thus the relation of Eq. (\ref{lewis}) is fulfilled.

To calculate the LR phase presented in Eq. (\ref{LRphase}), one is led to
notice that $|\vec{\alpha} (t)|=1$ and $I(t)$ of Eq. (\ref{invar}) can be
written as $I(t)=-G(t)J_zG^{\dagger }(t)$, in which $G(t)=e^{i\varphi
J_z}e^{i\theta (t)J_y}$ accounts for a canonical transformation with $\theta
(t)=\arccos \frac{-\nu t}{\sqrt{1+\nu ^2t^2}}$ and $\varphi =-\arcsin \frac
\nu \eta $. The eigenvector of $I(t)$ is then obtained as $|\phi
_m(t)\rangle =G(t)|m\rangle $, in which $|m\rangle $ $(m=-j,-j+1,\cdots j)$
represents the eigenstate of $J_z$. With these notations, the two terms
contained in the kernel of the integral of Eq. (\ref{LRphase}) can be worked
out straightforwardly. It happens that the first term $\langle \phi
_m(t)|i\partial _t|\phi _m(t)\rangle $, which denotes a nonadiabatic
counterpart of the geometric connection of the adiabatic evolution, always
vanishes in the present system. The second term $\langle \phi
_m(t)|H(t)|\phi _m(t)\rangle $ identifies the diabatic energy levels \cite{note2}
of the system and is shown to be
\begin{equation}
E_m(t)\equiv\langle \phi _m(t)|H(t)|\phi _m(t)\rangle =\frac{-m\eta \kappa }{\sqrt{
1+\nu ^2t^2}}.  \label{diab}
\end{equation}
As $\kappa\rightarrow 1$, they recover the adiabatic levels
$E_m^{ad}(t)=-m\eta/\sqrt{1+\nu^2 t^2}$. We illustrate both
$E_m^{ad}(t)$ and $E_m(t)$ for the $j=\frac 12$ case in Fig. 1(b).

The rigorous dynamical solution achieved above is applicable to the general
angular-momentum system with an arbitrary azimuthal quantum number $j$. It
indicates a significant difference from that of the original LZ model since
the exact LZ formula of the latter model, which has been achieved as an
asymptotical result of the Weber's parabolic cylinder functions \cite{LZ2},
applies only to the two-level system. Moreover, it is worthy to stress
that the demonstration of the overall dynamical invariant for the above model
is highly nontrivial as the original model does not possess such an
invariant \cite{wxq}. As will be shown in the below, it implies
that the survival
probability of the adiabatic state in this model, albeit the existence of
intermediate transitions associated with nonadiabatic effects, tends
asymptotically to the unit for the overall evolution. It suggests that the
nonadiabatic evolution of the model can serve as a protocol for
complete population transfer.

\section{Dynamical transitions in the two-level and multi-level systems}

\subsection{Protocols for nonadiabatic population transfer}
Following the expression of Eq. (\ref{invar}), $I(t)$ will evolve from $-J_z$
to $J_z$ along a geodesic curve in the Bloch space during the overall
evolution $t\in(-\infty, \infty )$. Since the eigenstates of $I(t)$ are
transported parallel without transitions, an initial eigenstate $|m\rangle$
then will evolve to the ending state $|-m\rangle $ at $t\rightarrow\infty$.
Therefore, up to a phase term, the generated dynamics yields complete
population transfer $|m\rangle\leftrightarrow |-m\rangle $ for the system
whatever the sweep process is adiabatic or nonadiabatic.

For an irreducible space spanned by the angular momentum operator with a
specific quantum number $j$, an explicit expression of the basis state
$|\phi_m(t)\rangle$ could be obtained via
\begin{eqnarray}
|\phi _m(t)\rangle &=&e^{i\varphi J_z}e^{i\theta J_y}|m\rangle  \nonumber \\
&=&\sum_{m^{\prime }}\mathcal{D}_{m^{\prime }m}^j(\theta )e^{im^{\prime
}\varphi }|m^{\prime }\rangle ,  \label{basis}
\end{eqnarray}
in which $\mathcal{D}_{m^{\prime }m}^j(\theta )\equiv \langle m^{\prime
}|e^{i\theta J_y}|m\rangle $ has explicit expression for the specified $j$
\cite{textbook} and the index $m^{\prime }$ of the summation is
taken over $-j,-j+1,\cdots ,j$. Specifically, for the two-level system with
$j=\frac 12$, one has
\begin{equation}
|\phi _{\pm }(t)\rangle =e^{\pm i\frac \varphi 2}\cos \frac{\theta (t)}2|\pm
\rangle \pm e^{\mp i\frac \varphi 2}\sin \frac{\theta (t)}2|\mp \rangle ,
\label{2basis}
\end{equation}
where we have used the notation ``$|\pm \rangle $" for $|\pm \frac 12\rangle
$. Accordingly, the phase-equipped dynamical basis $|\psi_{\pm}(t)\rangle$
is obtained straightforwardly and the process of the population transfer is
then characterized as
\begin{eqnarray}
|\psi _{+}(-\infty )\rangle &=&|+\rangle \rightarrow |\psi _{+}(\infty
)\rangle =e^{i\beta _{+}}|-\rangle ,  \nonumber \\
|\psi _{-}(-\infty )\rangle &=&|-\rangle \rightarrow |\psi _{-}(\infty
)\rangle =-e^{i\beta _{-}}|+\rangle ,  \label{transfer}
\end{eqnarray}
in which $\beta _{\pm }=\Phi _{\pm }(\infty )\mp \varphi $ with
\begin{equation}
\Phi _{\pm }(\infty )=\pm \frac 12\eta \kappa \int_{-\infty }^\infty (1+\nu
^2t^2)^{-\frac 12}dt.  \label{phase2}
\end{equation}

The driving field in the present scheme has finite intensity and tends
to zero asymptotically as $t\rightarrow \pm \infty $. In the practical
scanning process the driving field should be pulsed in a finite time duration
with truncation, that is, $t\in[-\tau_c,\tau_c]$. It turns out that
the cutoff of the pulse results in very limited influence on the transition
probability.
For simplicity, let us take the above two-level case as an example. The
transition probability induced by the sweep over the period
$t\in[-\tau_c,\tau_c]$ is defined by
$P\equiv|\langle -|U(\tau_c,-\tau_c)|+\rangle |^2$
(or equally by $|\langle +|U(\tau_c,-\tau _c)|-\rangle |^2$)
in which $U(\tau_c,-\tau _c)$ accounts for the generated evolution
operator
\begin{equation}
U(\tau _c,-\tau _c)=\sum_{\pm }e^{i\Phi _{\pm }(\tau _c,-\tau _c)}|\phi
_{\pm }(\tau _c)\rangle \langle \phi _{\pm }(-\tau _c)|.  \label{evolu}
\end{equation}
A straightforward calculation yields that
\begin{equation}
P=1-(1+\nu ^2\tau _c^2)^{-1}\cos ^2\frac{\Phi _{+}(\tau _c,-\tau _c)}2.
\label{prob}
\end{equation}
Loss of the fidelity of the population transfer, defined by
$P_\delta \equiv 1-P$ here, is limited by the factor $(1+\nu ^2\tau _c^2)^{-1}$.
It is seen that as $\nu \tau _c\geq10\pi $, the population transfer is realized
with a high fidelity: $P_\delta \lesssim 10^{-3}$.

\begin{figure}[t]
\includegraphics[width=0.9\columnwidth]{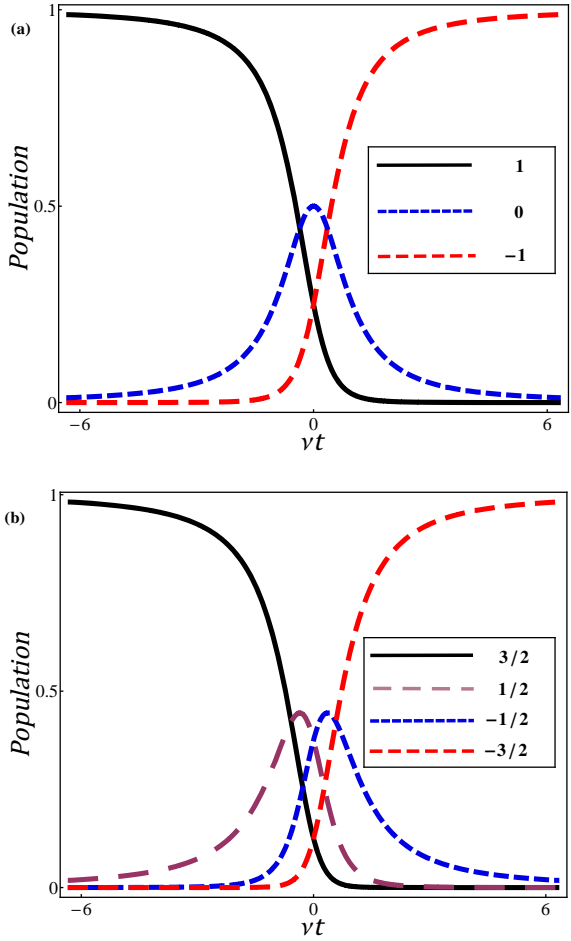}
\caption{Nonadiabatic population transfer along the time evolution in the
multi-level systems in which the parameters are set as $\nu/\eta=0.8$ and
$\kappa=0.6$.
(a) The three-level system with $j=1$ and the initial state is in $|1\rangle$.
The maximal population in the intermediate state $|0\rangle$ is obtained as
$p=\frac 12$ at $t=0$ . (b) The four-level system with $j=\frac 32$ and the
initial state is in $|\frac{3}{2}\rangle$. The maximal populations on the
states $|\frac 12\rangle$ and $|-\frac 12\rangle$ are obtained as $p=\frac{4}{9}$
at $\nu t=-\frac{1}{2\sqrt{2}}$ and $\frac{1}{2\sqrt{2}}$, respectively.}
\end{figure}

\subsection{Nonadiabaticity-induced transitions and survival probabilities of
the adiabatic states in the multi-level systems}
Dynamic control of the multi-channel nonadiabatic process is usually
a more challenging task and there have been extensive studies \cite{multi1,multi2,multi3}
on that of the multi-state version of the Landau-Zener model.
Intriguingly, the model we proposed above applies directly to the multi-level system
and the multi-channel dynamical transitions can be manifested by
exploring the model with high quantum number $j$. For the cases of $j=1$
and $j=\frac 32$, the corresponding representative matrices $\mathcal{D}
^j(\theta) $ are expressed explicitly as
\begin{equation}
\mathcal{D}^1(\theta )=\left(
\begin{array}{lll}
\cos ^2\frac \theta 2 & \frac{\sin \theta }{\sqrt{2}} & \sin ^2\frac \theta 2
\\
-\frac{\sin \theta }{\sqrt{2}} & \cos \theta & \frac{\sin \theta }{\sqrt{2}}
\\
\sin ^2\frac \theta 2 & -\frac{\sin \theta }{\sqrt{2}} & \cos ^2\frac \theta
2
\end{array}
\right)  \label{D3}
\end{equation}
and
\begin{equation}
\mathcal{D}^{\frac 32}(\theta )=\left(
\begin{array}{llll}
\cos ^3\frac \theta 2 & d_{12} & d_{13} & \sin ^3\frac \theta 2
\\
-d_{12} & d_{22} & d_{23} & d_{13} \\
d_{13} & -d_{23} & d_{22} & d_{12} \\
-\sin ^3\frac \theta 2 & d_{13} & -d_{12} & \cos ^3\frac \theta 2
\end{array}
\right),  \label{D4}
\end{equation}
where $d_{12}=\frac{\sqrt{3}}2\cos \frac \theta 2\sin \theta $,
$d_{13}=\frac{\sqrt{3}}2\sin \frac \theta 2\sin \theta $,
$d_{22}=3\cos^3\frac \theta 2-2\cos \frac \theta 2$
and $d_{23}=2\sin \frac \theta 2-3\sin^3\frac \theta 2$.
The processes of
the population transfer in these two cases are depicted in Fig. 2, in which
the initial states are taken to be $|1\rangle$ and $|\frac 32\rangle$,
respectively.

To characterize further the nonadiabatic effects in the dynamical evolution,
we evaluate the matrix of the transition probability: $T_{mn}=|\langle
\psi _m^{ad}(t)|\psi _{n}(t)\rangle |^2$, in which $|\psi
_m^{ad}(t)\rangle$ stands for the instantaneous adiabatic eigenvector of the
Hamiltonian (\ref{hamil1}). The diagonal elements of the matrix $T$
represent the survival probabilities of the adiabatic basis states and the
off-diagonal ones describe unambiguously the nonadiabaticity-induced transitions
between these adiabatic states along the evolution.
It is recognized that the basis set $|\psi_m(t)\rangle $,
that are identical to $|\psi_m^{ad}(t)\rangle $ at the initial time
$t\rightarrow -\infty$, will exhibit intermediate transitions during the
evolution. However, as $|\psi_m(t)\rangle $ will recover
$|\psi_m^{ad}(t)\rangle $ (up to a phase factor) eventually at $t\rightarrow \infty$,
the desired population transfer is not destroyed by these
nonadiabaticity-induced transitions.
In Fig. 3 we illustrate in detail these phenomena for the model with $j=1$
and $j=\frac 32$.

\begin{figure}[t]
\includegraphics[width=0.9\columnwidth]{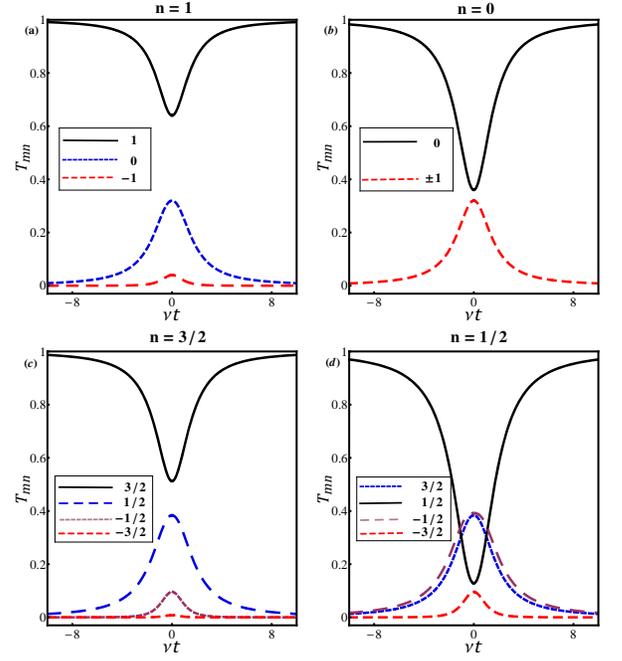}
\caption{Survival probability of the adiabatic states
(the diagonal elements $T_{nn}$) and
the nonadiabaticity-induced transition
(the off-diagonal elements $T_{mn}$ with $m\neq n$)
in the dynamical evolution.
The parameters are set as $\nu/\eta=0.8$ and $\kappa=0.6$.
(a) The $j = 1$ model with an initial state $|1\rangle$ $(n=1)$. (b) The
$j = 1$ model with an initial state $|0\rangle$ in which the intermediate
transitions to $|\psi_{1}^{ad}(t)\rangle$ and to
$|\psi_{-1}^{ad}(t)\rangle$ have an equal
probability. (c) The $j =\frac 32$ model in which the initial
state is in $|\frac 32\rangle$ $(n=\frac 32)$ and all the elements $T_{mn}$
$(m=\pm\frac 32,\pm\frac 12)$ are characterized.
(d) The $j =\frac 32$ model in which the initial
is in $|\frac 12\rangle$ $(n=\frac 12)$.  }
\end{figure}

\section{Noise effects in the presence of dissipation}

In realistic systems the noise due to the surrounding environment is
inevitable. The influence of the system-bath coupling to the transition
probability for the original LZ model has ever been studied in various
background \cite{dissp1,dissp2,dissp3,dissp4}. In the following we shall
investigate the noise effect on the dynamics for the present modulated LZ
model. Typically, we focus on the two-level system and estimate
the population transfer in the presence of the spin flip noise which can arise
as the interaction of the spin system with its fermionic reservoir
is involved \cite{fermion1,fermion2}. Within the Markovian regime,
the evolution of the system is described by the master equation
\begin{equation}
\frac{\partial \rho (t)}{\partial t}=-i[H(t),\rho (t)]-\sum_i\frac{\gamma _i}
2[J_i,[J_i,\rho (t)]],  \label{master}
\end{equation}
where $\gamma_i$ $(i=x,y,z)$ accounts for the damping rate of the corresponding
spin flip process. Since we only consider the $j=\frac 12$ case,
it is convenient to introduce the Bloch
vector $(\rho _x,\rho _y,\rho _z)$ to describe the elements of the density
operator, that is, $\rho _x=\rho _{+-}+\rho _{-+}$, $\rho _y=-i(\rho
_{+-}-\rho _{-+})$ and $\rho _z=\rho _{++}-\rho _{--}$.
According to Eq. (\ref{master}), one obtains that these components satisfy
\begin{equation}
\frac \partial {\partial t}\left(
\begin{array}{l}
\rho _x \\
\rho _y \\
\rho _z
\end{array}
\right) =-\left(
\begin{array}{lll}
\frac{\gamma _y+\gamma _z}2 & \Omega _z & 0 \\
-\Omega _z & \frac{\gamma _x+\gamma _z}2 & \Omega _x \\
0 & -\Omega _x & \frac{\gamma _x+\gamma _y}2
\end{array}
\right) \left(
\begin{array}{l}
\rho _x \\
\rho _y \\
\rho _z
\end{array}
\right) .  \label{master2}
\end{equation}

\begin{figure}[t]
\includegraphics[width=0.9\columnwidth]{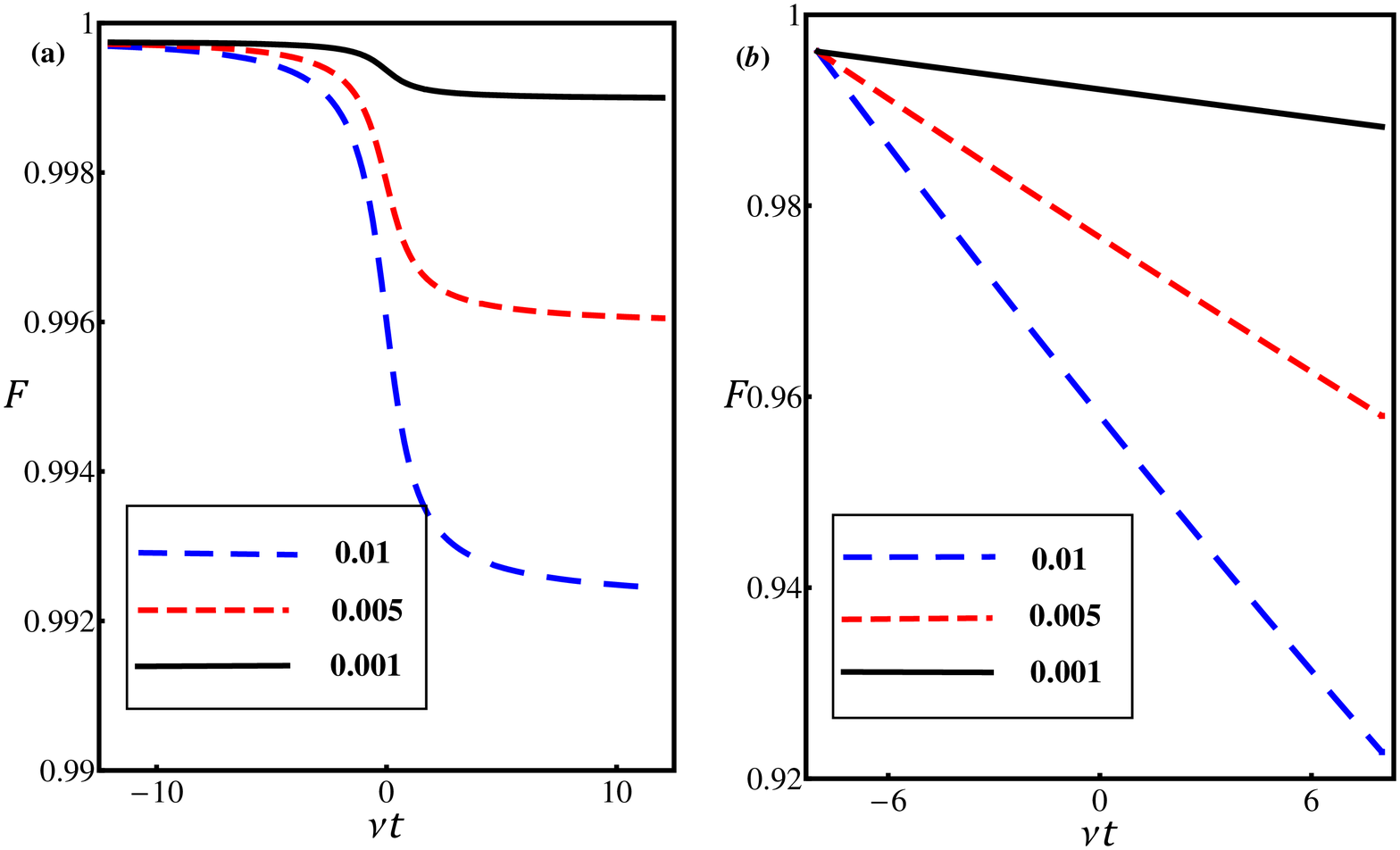}
\caption{Time evolving of the fidelity $F(t)$ of the driving protocol
($\kappa=0.6$ and $\nu/\eta=0.8$) in the presence of noise effects.
(a) Pure phase damping process in which $\gamma_{x,y}=0$
and $\gamma_z/\nu=0.01,0.005$ and $0.001$, respectively. The duration
of the pulse is chosen to be $\nu \tau_c=10\pi$ and the final fidelity
is achieved as $F(\tau_c)\approx 0.992, 0.996$ and $0.999$,
respectively. (b) The random spin flip process with
$\gamma_{x,y,z}=\gamma$. The duration of the pulse is chosen to be
$\nu \tau_c=8$ and the fidelities $F(\tau_c)\approx 0.923,0.958$ and $0.988$
are obtained for $\gamma/\nu=0.01,0.005$ and $0.001$, respectively.}
\end{figure}

For the situation $\gamma_x=\gamma_y=0$, the above loss mechanism
accounts for the pure phase
damping process which does not lead to direct transitions between the two levels
$|\pm\rangle $.
Nevertheless, as the dephasing process will alter the trajectory of the dynamical
evolution generated by $H(t)$, it will result in imperfect effect on the desired
population transfer. To characterize the influence of the
noise on the dynamical evolution, the central task is to compute the
fidelity $F(t)=|\langle \phi _+(t)|\rho(t)|\phi _+(t)\rangle|$, in which
$|\phi _+(t)\rangle$ is given explicitly in Eq. (\ref{2basis}). It is recognized
that $F(t)$ describes the overlap between the actual time-evolving state and
the target dynamical basis $|\phi_+(t)\rangle$.
Starting from an initial state $|+\rangle$, we solve numerically
the set of equations ($\ref{master2}$) for both the dephasing
process and the random spin flip process with $\gamma_{x,y,z}=\gamma$.
For the dephasing process, it happens that reduction of the fidelity
mainly occurs in the vicinity of the point $t=0$. We have chosen the
time duration $\nu\tau_c=10\pi$ and
the influence of the cutoff error is negligible. The result shows that
the driving protocol is insensitive to the dephasing and a fidelity
higher than $0.99$ is obtained even the ratio $\gamma/\nu\gtrsim 10^{-2}$.
On the other hand, the spin flip noise with the homogeneous damping rate
will exert detrimental effects on the desired state transfer continuously
over the whole time evolution.
In our calculation we choose $\nu\tau_c=8$. Besides the noise effect,
the dramatic truncation of the scanning process here has slight influence on
the population transfer. The yielded results about the time evolving of
the fidelity $F(t)$ are illustrated in Fig. 4, in which different
values of the ratio $\gamma/\nu$ are assumed.

\section{Conclusion}

We have investigated the exact dynamics of a modulated LZ model
and exploited it as a design for nonadiabatic quantum control.
Differing from the original LZ model, we have shown that this modulated model
possesses an analytical dynamical invariant over the whole time
domain and the generated dynamics is fully solvable analytically.
While serving as a protocol for population transfer,
the model is shown to possess the following distinct advantages:
1) nonadiabatic dynamics generated by the model
itself can realize complete population transfer;
2) the protocol uses only finite driving fields which avoids
the nonrealistic ingredient assuming infinite driving
in the original LZ and also other analogous protocols.
Furthermore, the scheme is applicable to the multi-level
systems which offers an unambiguous scenario to manifest
the multi-channel transitions induced by the nonadiabatic effects
in the state transfer process.

As the noise due to the dissipative environment
will lead to detrimental effects on the desired
control process, we have also investigated the loss of the fidelity
for the protocol when the system is subjected to the dissipation.
The numerical calculations reveal that the protocol is not
sensitive to the pure phase damping noise.
On the other hand, to obtain high-fidelity population transfer
in the presence of the spin flip noise with a homogeneous damping rate $\gamma$,
our calculation shows that a requirement of the scanning rate of the protocol,
$\nu/\gamma\gtrsim 10^3$, should be satisfied in general.
Suppose that the coherence time $\gamma ^{-1}$ is of an order $\sim10^2~\mu\rm{s}$
(which is achievable for the electron spin of the nitrogen-vacancy center
in diamond \cite{exp1}), then approximative evaluation yields that
the sweep frequency should be $\nu\gtrsim 10~\rm{MHz}$ and
the time duration of the pulse $\tau_c\sim \pi~\mu\rm{s}$.
By taking $\kappa=0.6$ (cf. Fig. 1)
one gets $\eta\sim12.5~\rm{MHz}$. Potential experimental implementation of the
protocol in physical systems is highly expected.


\begin{references}

\bibitem{nonadt} H. Nakamura, {\it Nonadiabatic Transition} (World Scientific,
Singapore, 2002).


\bibitem{brumer} P. W. Brumer and M. Shapiro, {\it Principles of the Quantum
Control of Molecular Processes} (Wiley-Interscience, New York, 2003).



\bibitem{LZ1} L.D. Landau, Phys. Z. Sowjetunion {\bf 2}, 46 (1932).

\bibitem{LZ2} C. Zener, Proc. R. Soc. A {\bf 137}, 696 (1932).


\bibitem{cd1} M. Demirplak and S. A. Rice, J. Phys. Chem. A {\bf 107}, 9937
(2003).

\bibitem{cd2} M. Demirplak and S. A. Rice, J. Phys. Chem. B {\bf 109}, 6838
(2005).

\bibitem{tranl} M. Berry, J. Phys. A: Math. Theor. {\bf 42}, 365303 (2009).

\bibitem{short} X. Chen, I. Lizuain, A. Ruschhaupt, D. Gu¡äery-Odelin,
and J.G. Muga, Phys. Rev. Lett. {\bf 105}, 123003 (2010).

\bibitem{barn1} E. Barnes and S. Das Sarma, Phys. Rev. Lett. {\bf 109},
060401 (2012).

\bibitem{barn2} E. Barnes, Phys. Rev. A {\bf 88}, 013818 (2013).

\bibitem{mess} A. Messina and H. Nakazato, J. Phys. A {\bf 47}, 445302 (2014).

\bibitem{exp1} J. Zhang, J.H. Shim, I. Niemeyer, T. Taniguchi, T. Teraji,
H. Abe, S. Onoda, T. Yamamoto, T. Ohshima, J. Isoya,
and D. Suter, Phys. Rev. Lett. {\bf 110}, 240501 (2013).

\bibitem{exp2} M.G. Bason, M. Viteau, N. Malossi, P. Huillery, E. Arimondo,
D. Ciampini, R. Fazio, V. Giovannetti, R. Mannella,
and O. Morsch, Nat. Phys. {\bf 8}, 147 (2012).

\bibitem{exp3} B.B. Zhou, A. Baksic, H. Ribeiro, et al.,
Nat. Phys. {\bf 13}, 330 (2017).

\bibitem{LZI1} W.D. Oliver, Y. Yu, J.C. Lee, K.K. Berggren, L.S. Levitov,
and T.P. Orlando, Science {\bf 310}, 1653 (2005).

\bibitem{LZI2} S.N. Shevchenko, S. Ashhab, and F. Nori,
Phys. Rep. {\bf 492}, 1 (2010).

\bibitem{LZI3} S. Gasparinetti, P. Solinas, and J.P. Pekola,
Phys. Rev. Lett. {\bf 107}, 207002 (2011).

\bibitem{LZI4} F. Forster, G. Petersen, S. Manus, P. H\"{a}nggi,
D. Schuh, W. Wegscheider, S. Kohler, and S. Ludwig,
Phys. Rev. Lett. {\bf 112}, 116803 (2014).

\bibitem{charge} A.M. Kuztetsov, {\it Charge Transfer in Physics, Chemistry,
and Biology} (Gordon and Breach, Reading, 1995).

\bibitem{chem1} C. Zhu and S.H. Lin, J. Chem. Phys. {\bf 107}, 2859 (1997).

\bibitem{chem2} A. Nitzan, {\it Chemical Dynamics in Condensed Phases}
(Oxford University Press, Oxford, 2006).

\bibitem{manip} L.F. Wei, J.R. Johansson, L.X. Cen, S. Ashhab,
and Franco Nori, Phys. Rev. Lett. {\bf 100}, 113601 (2008).

\bibitem {triq} G. Sun, X. Wen, B. Mao, J. Chen, Y. Yu, P. Wu,
and S. Han, Nat. Commun. {\bf 1}, 51 (2010).

\bibitem{you} J.Q. You and F. Nori, Nature (London) {\bf 474}, 589 (2011).

\bibitem{qtrit} M.B. Kenmoe, L.C. Fai, Phys. Rev. B {\bf 94}, 125101 (2016)

\bibitem{rosen} N. Rosen and C. Zener, Phys. Rev. {\bf 40}, 502 (1932).

\bibitem{allen} L. Allen and J. H. Eberly, {\it Optical Resonance and Two-Level
Atoms} (Dover, New York, 1975).

\bibitem{demkov} Y.N. Demkov and M. Kunike, Vestn. Leningr. Univ., Ser. 4, Fiz.
Khim. {\bf 16}, 39 (1969).

\bibitem{bamb} A. Bambini and P. R. Berman, Phys. Rev. A {\bf 23}, 2496 (1981).


\bibitem{tan} G. Yang, W. Li, and L.-X. Cen, arXiv: 1608.00735.


\bibitem{lewis1} H.R. Lewis Jr., Phys. Rev. Lett. {\bf 18}, 510 (1967).

\bibitem{lewis2} H.R. Lewis Jr. and W.B. Riesenfeld, J. Math. Phys. {\bf 10}, 1458 (1969).

\bibitem{wang} S.J. Wang, Phys. Rev. A {\bf 42}, 5107 (1990).

\bibitem{HQC} L.-X. Cen, X.Q. Li, Y.J. Yan, H.Z. Zheng, and S.J. Wang,
Phys. Rev. Lett. {\bf 90}, 147902 (2003).

\bibitem{Dang} T.T. Nguyen-Dang, E. Sinelnikov, A. Keller, and O. Atabek,
Phys. Rev. A {\bf 76}, 052118 (2007).

\bibitem{note2} The term ``diabatic" has different usage in some
other literatures, e.g., in the community of the multi-level
Laudau-Zener model (see Refs. \cite{multi1,multi2,multi3}).

\bibitem{wxq} X.Q. Wang and L.-X. Cen, Phys. Lett. A {\bf 375}, 2220 (2011).

\bibitem{textbook} E.P. Winger, {\it Group Theory and Its Application to the Quantum
Mechanics of Atomic Spectra} (Academic Press, New York, 1959).

\bibitem{multi1} Y.N. Demkov and V.N. Ostrovsky, Phys. Rev. A {\bf 61},
032705 (2000).

\bibitem{multi2} N.A. Sinitsyn, J. Phys. A {\bf 48}, 195305 (2015).

\bibitem{multi3} N.A. Sinitsyn, J. Lin, V.Y. Chernyak, Phys. Rev. A {\bf 95},
012140 (2017).

\bibitem{dissp1} M. Wubs, K. Saito, S. Kohler, P. H\"{a}nggi, and Y. Kayanuma,
Phys. Rev. Lett. {\bf 97}, 200404 (2006).

\bibitem{dissp2} S. Ashhab, Phys. Rev. A {\bf 90}, 062120 (2014).

\bibitem{dissp3} S. Javanbakht, P. Nalbach, and M. Thorwart, Phys. Rev. A
{\bf 91}, 052103 (2015).

\bibitem{dissp4} Z. Sun, L. Zhou, G. Xiao, D. Poletti, and J. Gong, Phys.
Rev. A {\bf 93}, 012121 (2016).

\bibitem{fermion1} N. Yamada, A. Sakuma, and H. Tsuchiura,
J. Appl. Phys. {\bf 101}, 09C110 (2007).

\bibitem{fermion2} M.W. Wu, J.H. Jiang, and M.Q. Weng, Phys. Rep. {\bf 493}, 61 (2010).

\end{references}
\end{document}